\RequirePackage{fix-cm}
\documentclass[smallextended]{svjour3}       
\smartqed  
\usepackage{graphicx}
\usepackage{multirow}
\usepackage{epsfig}
\journalname{}

\title{GHZ paradoxes based on an even number of qubits}

\author{Mordecai Waegell and P.K. Aravind }

\authorrunning{M.Waegell, P.K. Aravind}

\institute{M.Waegell, P.K. Aravind \at
Physics Department, Worcester Polytechnic Institute, Worcester, MA 01609, U.S.A.\\
\email{caiw@wpi.edu, paravind@wpi.edu}}

\date{\today}

\begin{document}
\maketitle
\begin{abstract}
GHZ paradoxes are presented for all even numbers of qubits from four up. They are obtained from proofs of the Kochen-Specker (KS) theorem by showing how the assumption of noncontextuality can be justified on the basis of locality. The nature of the entangled states involved in our paradoxes is discussed. Some multiqubit proofs of the KS theorem are also presented in the form of diagrams from which they are visually obvious. The implications of our results are discussed.
\end{abstract}\\\
\section{\label{sec:Intro}Introduction}
After Greenberger, Horne and Zeilinger \cite{GHZ} gave their proof of Bell's theorem without inequalities \cite{Redhead}, there have been many refinements and extensions of their results \cite{Mermin1,Mermin2,Pagonis,DiV,Zuk,Cerf,Lee,Tang,Cabello2001,Aravind2002}. Several authors have proposed GHZ paradoxes (as we will term proofs of Bell's theorem without inequalities or probabilities) based on qudits (i.e., $d$-dimensional quantum systems): Ref. \cite{Zuk} constructs paradoxes based on a maximally entangled state of $d$ qudits shared among $d$ observers; Ref. \cite{Cerf} presents a variety of paradoxes based on an arbitrary number of qudits and also clarifies what it means for a paradox to be genuinely multipartite and multidimensional; and Ref. \cite{Tang} discusses paradoxes based on graph states of qudits. Despite this progress, it is surprising that GHZ paradoxes have not been formulated for an arbitrary even number of qubits (by contrast, Mermin \cite{Mermin1} presented a proof for three qubits and DiVincenzo and Peres \cite{DiV} one for five, from which a generalization to all odd numbers is easily achieved \cite{Aravind2002,Tang}). It is the purpose of this Letter to fill this gap by presenting GHZ paradoxes for all even numbers of qubits from four up \cite{Note1}. Our paradoxes are interesting because they involve  entangled states that are distinct from GHZ, W, cluster or graph states. The state $\Psi_{4}$ involved in our four-qubit paradox has been used earlier by Yeo and Chua \cite{Yeo} to discuss teleportation and dense coding, but the fact that it can be used to demonstrate a GHZ paradox does not seem to have been noticed. It is also interesting that $\Psi_{4}$ is the lowest member of an infinite class of states that gives rise to GHZ paradoxes and whose members can be expressed as sums of products of Bell states. The second purpose of this Letter is to draw attention to a large number of multiqubit proofs of the Kochen-Specker (KS) theorem based on the observables of the N-qubit Pauli group. We will discuss the implications of our results after we have first presented them.\\
\section{\label{sec:2}GHZ paradoxes for an even number of qubits}
\begin{table}[ht]
\begin{minipage}[b]{0.5\linewidth}\centering
\begin{tabular}{|c c c c|} 
\hline 
Z & Z & Z & Z \\
X & X & Z & Z \\
Z & X & X & I \\
X & Z & I & X \\
I & I & X & X \\
\hline
\end{tabular}
\end{minipage}
\hspace{0.2cm}
\begin{minipage}[b]{0.5\linewidth}
\centering
\begin{tabular}{|c c c c c c c c|} 
\hline 
Z & Z & Z & Z & Z & Z & $\cdots$ & Z \\
X & X & Z & Z & Z & Z & $\cdots$ & Z \\
Z & X & X & I & I & I & $\cdots$ & I \\
X & Z & I & X & I & I & $\cdots$ & I \\
I & I & I & X & X & I & $\cdots$ & I \\
I & I & I & I & X & X & $\ddots$ & $\vdots$ \\
$\vdots$ & $\vdots$ & $\vdots$ & $\vdots$ & $\ddots$ & $\ddots$ & $\ddots$ & I \\
I & I & I & I & $\cdots$ & I & X & X \\
I & I & X & I & $\cdots$ & I & I & X \\
\hline
\end{tabular}
\end{minipage}
\caption{Left: Five mutually commuting observables of a four-qubit system, arranged horizontally, with their corresponding qubits vertically aligned.  Right: $2N+1$ mutually commuting observables of a $2N$-qubit system, for $N \geq 3$, in the same format as at left.}
\label{Star4N} 
\end{table}

Table 1, left, shows five mutually commuting observables of a four-qubit system arranged in the form of a $5 \times 4$ array, with the rows representing the observables and the columns their corresponding qubits ($X,Y,Z$ and $I$ are the Pauli and identity operators of the qubits). A noncontextual hidden variables theory (NHVT) is required to assign the value $+1$ or $-1$ to each of the four-qubit observables, as well as all the single qubit observables of which they are made up, in such a way that all the operator relations satisfied by the observables are mirrored in algebraic relations satisfied by their values. This requires that the value assigned to any four-qubit observable be equal to the product of the values assigned to its single-qubit constituents (for example, $v(ZXXI) = v(Z_{1})v(X_{2})v(X_{3})$, where subscripts have been used on the right to indicate the qubits involved) and also that $v(ZZZZ)v(ZXXI)v(XXZZ)v(XZIX)v(IIXX) = -1$ (since the product of these observables is the negative of the identity operator). However, if one reexpresses the left side of the last equation in terms of the values of the single-qubit observables, one finds that each value occurs twice, implying that the left side is $+1$. But this leads to the contradictory equation $+1 = -1$, which shows that NHVTs are untenable and proves the KS theorem.\\

We now show how the above KS proof can be converted into a GHZ paradox. Let $\Psi_{4}$ be the simultaneous eigenstate of the four-qubit observables of Table 1, left, with eigenvalue $+1$ for the first four observables and $-1$ for the last. Suppose a source repeatedly emits four qubits in the state $\Psi_{4}$, sending one qubit to each of the observers A,B,C and D who are all at spacelike separations from one another. Suppose each observer randomly measures $X$, $Z$ or $I$ on his/her qubit in each of the runs ($I$ means they simply do nothing) and they get together later to analyze their results, keeping only the runs in which their measurements make up the single-qubit observables in one of the rows of Table 1, left. We now show that all the single qubit observables in Table 1, left, are ``elements of reality", i.e., they have values that can be determined without disturbing the qubits in any way. This is so because, in any run, the value of any one of the single-qubit observables can be deduced from the observed values of all the others if one uses the fact that the product of their values is fixed (and equal to the eigenvalue of the corresponding four-qubit observable). Locality also guarantees that this value is independent of which alternative sets of measurements are carried out on the other qubits, or that this value is noncontextual. The earlier KS argument then shows that it is impossible to assign values to all the single qubit observables in such a way that all the four-qubit observables have their stated (eigen)values, and one gets a full fledged GHZ paradox.\\

Table 1, right, shows a generalization of the four-qubit paradox to one for $2N$ qubits, for $N \geq 3$. As before, the KS proof can be converted into a GHZ paradox by making use of any joint eigenstate of the $2N$-qubit observables. We have checked, with the aid of a computer program, that the proofs for $4,6,8$ and $10$ qubits are genuine multipartite proofs \cite{Cerf} in the sense that the elimination of any columns (qubits) and/or rows (multiqubit observables) from Table 1, right, does not leave a valid proof. We conjecture that this property holds for all higher (even) numbers of qubits, but do not have a rigorous proof of it.\\

We next discuss the structure of the joint eigenstates that occur in our paradoxes. The state $\Psi_{4}$ can be expressed in terms of the standard Bell states $\Phi^{\pm}$ and $\Psi^{\pm}$ as $\Psi_{4} = \Phi^{+}_{12}\Phi^{-}_{34} - \Psi^{-}_{12}\Psi^{-}_{34}$, where the subscripts label the qubits (which are numbered in ascending order from left to right in all our observables). State $\Psi_{4}$ has the property that a measurement on any two qubits in the computational basis leaves the other two qubits in a Bell state. Briegel and Raussendorf \cite{Briegel} discussed a state with this property and Yeo and Chua \cite{Yeo} used essentially the state $\Psi_{4}$ to discuss teleportation and dense coding of a two-qubit system. The joint eigenstate of our six-qubit paradox, with eigenvalue $-1$ for the last observable and $+1$ for all the others, is

\begin{eqnarray}\nonumber
\Psi_{6} = \Phi^{+}_{12}(\Phi^{-}_{34}\Phi^{+}_{56} + \Psi^{-}_{34}\Psi^{+}_{56}) - \Psi^{-}_{12}(\Phi^{-}_{34}\Psi^{+}_{56} + \Psi^{-}_{34}\Phi^{+}_{56}) \\
\nonumber
= (00)_{13}(\Phi^{-}_{24}\Phi^{+}_{56} + \Psi^{-}_{24}\Psi^{+}_{56})- (01)_{13}(\Psi^{-}_{24}\Phi^{+}_{56} + \Phi^{-}_{24}\Psi^{+}_{56}) \\
\nonumber
+ (10)_{13}(\Psi^{+}_{24}\Phi^{+}_{56} + \Phi^{+}_{24}\Psi^{+}_{56}) - (11)_{13}(\Phi^{+}_{24}\Phi^{+}_{56} + \Psi^{+}_{24}\Psi^{+}_{56})
\end{eqnarray}
The above decompositions, and other similar ones, show that a measurement on any two qubits (in the computational basis) leaves the other four in a state similar to $\Psi_{4}$. State $\Psi_{6}$ is closely related to the six-qubit state of Borras et al \cite{Borras} that has been used to discuss teleportation and state sharing in two- and three-qubit systems \cite{Choudhury}. The eigenstate of our eight-qubit paradox can be expressed as a sum of eight terms, each of which is a product of four Bell states, and has the property that a measurement on any two qubits leaves the other six in a state similar to $\Psi_{6}$. We expect this hierarchical structure to persist for all the higher eigenstates.\\

\begin{table}[ht]
\begin{minipage}[b]{0.5\linewidth}\centering
\begin{tabular}{|c c c c c c|} 
\hline 
Z & Z & Z & Z & Z & Z \\
X & X & X & X & X & X \\
Z & X & Z & X & I & I \\
X & Z & I & I & Z & X \\
I & I & X & Z & X & Z \\
\hline
\end{tabular}
\end{minipage}
\hspace{0.2cm}
\begin{minipage}[b]{0.5\linewidth}
\centering
\begin{tabular}{|c c c c c c c c|} 
\hline 
Z&Z&Z&Z&Z&Z&Z&I \\
X&X&X&X&X&X&I&Z \\
Z&X&Z&X&I&I&Z&Z \\
X&Z&I&I&I&I&X&X \\
I&I&X&Z&Z&X&I&I \\
I&I&I&I&X&Z&X&X \\
\hline
\end{tabular}
\end{minipage}
\caption{GHZ paradoxes based on a set of six-qubit observables (left) and a set of eight-qubit observables (right), in the same format as Table 1.}
\label{6qubit} 
\end{table}

The only GHZ paradox that can be constructed for a four-qubit system is the one shown in Table 1, up to permutations of the columns and/or the operators $X,Y$ and $Z$ \cite{Note2}. For $N \geq 5$ it is possible to construct paradoxes that are more economical than the ones of Table 1 in that they involve less than $2N+1$ mutually commuting $2N$-qubit observables. Table 2 shows two such examples, one for six qubits and the other for eight that involve only five and six observables, respectively, rather than the seven and nine of Table 1.  A study of the entangled states associated with these paradoxes, and their relationship to the earlier ones, would be of interest.\\
\section{\label{sec:3} Multi-qubit proofs of the KS theorem}
We next present several multiqubit proofs of the KS theorem, which are similar to (but more involved than) the two- and three-qubit proofs that have been presented earlier \cite{Mermin1,Waegell2012}. Figure 1 shows two different proofs based on four qubits. The observables are placed within circles and mutually commuting sets of observables are joined by lines, with a thin or a thick line being used for a set whose product is {\bf +I} or {\bf $-$I} ({\bf I} is the identity operator in the space of all the qubits, and all the commuting sets in our diagrams are of one of these two types). It is easy to verify that the diagrams prove the KS theorem by noting that they have an odd number of thick lines and that each observable lies on an even number of lines. However these KS proofs cannot be converted into GHZ paradoxes because each involves more than one set of commuting multiqubit observables, and using an eigenstate of one to set up a paradox does not allow the observables of the others to be established as elements of reality. The kite and wheel diagrams of Fig.1 are both members of infinite families of proofs that stretch upwards for all numbers of qubits. The higher members of the kite family have more observables strung out along the tail of the kite, but the wheel family is more involved, with the diagrams for odd and even numbers of qubits having a different structure. We have discovered a large number of other KS proofs for systems of up to ten qubits using a specially designed computer program \cite{WaegellUn}; all the proofs can be depicted in the form of diagrams like Fig.1 and all are irreducible \cite{Cerf} in the sense that they fail if even a single qubit or observable is dropped from the set.

\begin{figure}
\centering
\begin{tabular}{cc}
\epsfig{file=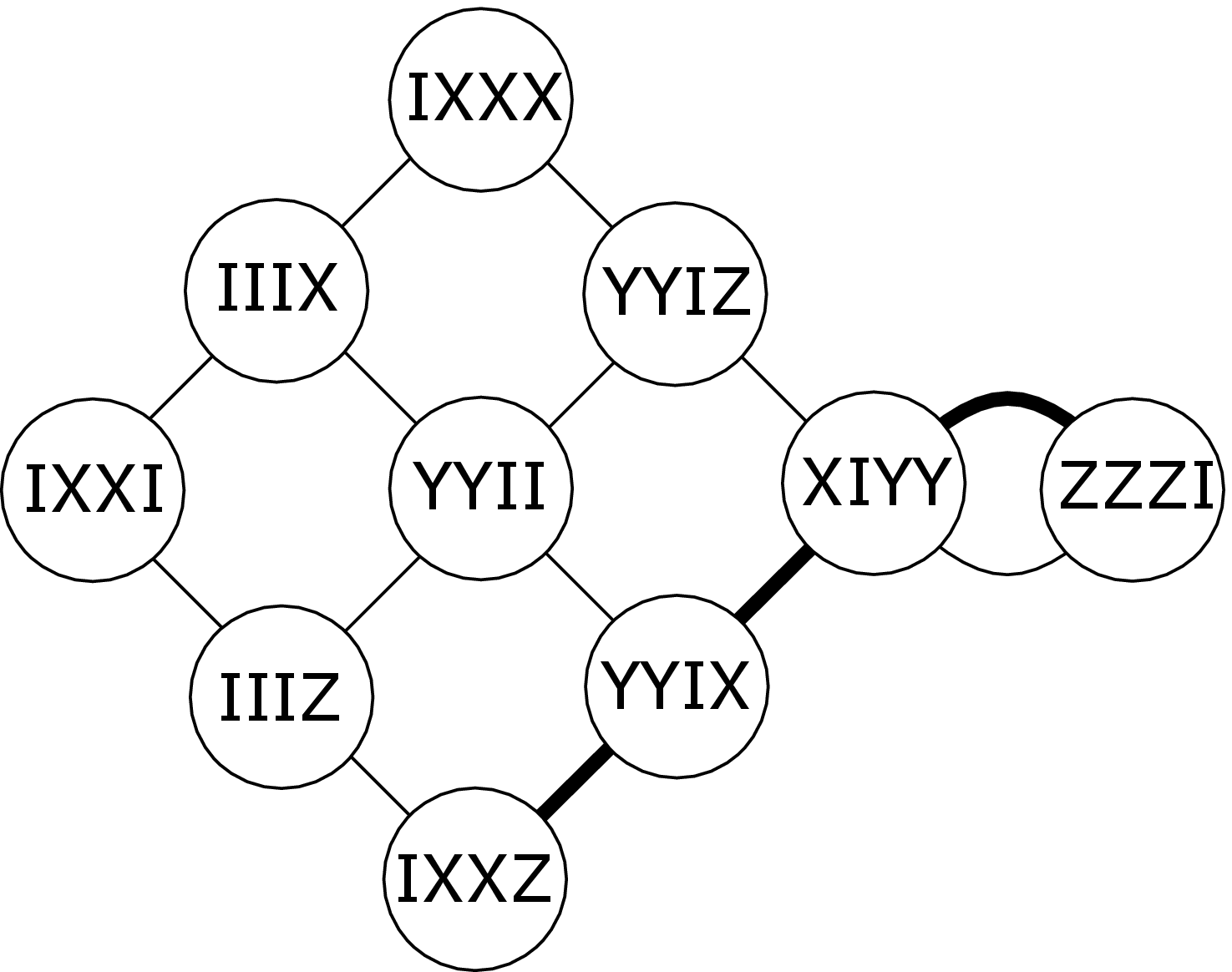,width=0.5\linewidth,clip=} &
\epsfig{file=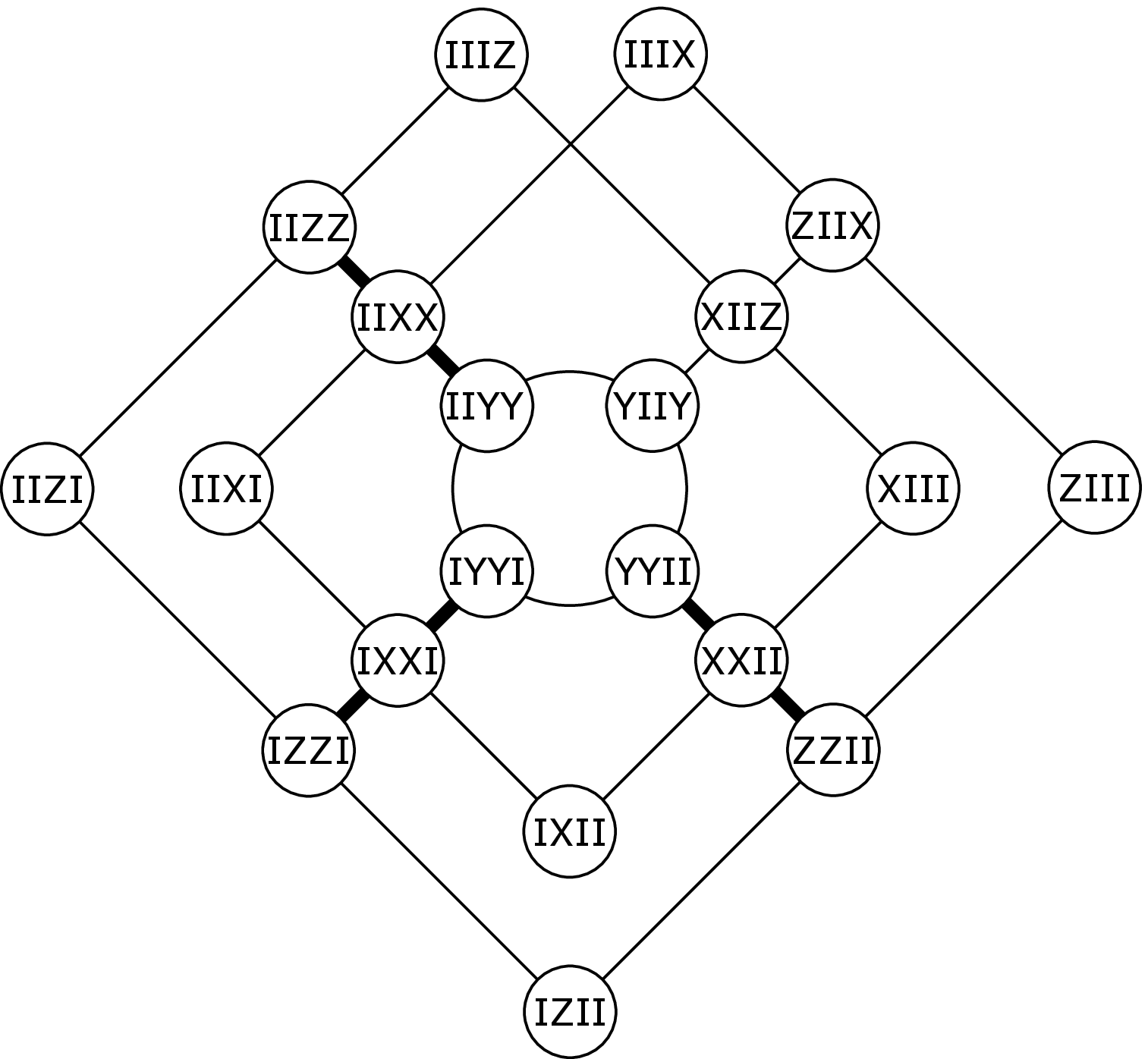,width=0.45\linewidth,clip=} \\
\end{tabular}
\caption{The four-qubit ``kite" (left) and four-qubit ``wheel" (right). The kite consists of two sets of four commuting observables (one of them is $IXXZ,YYIX,XIYY$ and $ZZZI$, joined by a thick line, and the other is $IXXX,YYIZ,XIYY$ and $ZZZI$, joined by a thin line). The wheel consists of only one set of four commuting observables, arranged around the thin circle at its center. The sets of three commuting observables in both diagrams are always connected by straight lines.}
\label{Fig.3}
\end{figure}
\section{\label{sec:4}Parity proofs of the KS theorem}
Our observables-based KS proofs (which we will refer to as ``KS systems", for short) can be used to obtain ``parity proofs" of the KS theorem based on sets of projectors and bases derived from them. A ``parity proof" (in a state space of dimension $\geq 4$) consists of a set of projectors, of possibly different ranks, that forms an odd number of bases (defined as a set of mutually orthogonal projectors that sums to the identity) in such a way that each projector occurs in an even number of the bases. Such a configuration proves the KS theorem because it is impossible for a NHVT to assign a 0 or a 1 to each of the projectors in such a way that the sum of the values in every basis is equal to the unity. The first parity proof, discovered by Cabello et al \cite{Cabello1996}, involved 18 rank-1 projectors and 9 bases in a state space of four dimensions. Since then thousands of other parity proofs have been discovered in four and eight dimensions \cite{Waegell2011,KP1995,Waegell2012}. We show here how parity proofs can be obtained in $2^{N}$ dimensions for $N \geq 4$. \\

To obtain all the parity proofs associated with a KS system, one begins by constructing the simultaneous eigenstates of all the sets of commuting observables in it. Instead of the eigenstates of the observables we will talk of the associated projectors, which are of rank-1 if the commuting set is complete and of higher rank if it is not. The projectors form two types of bases, that we will term ``pure" and ``hybrid". The pure bases are just the ones that arise as the simultaneous eigenstates of the commuting sets of observables, while the hybrids are made up of mixtures of projectors from different pure bases. We will term the set of all bases, both pure and hybrid, as the ``basis table" of the KS system. A remarkable feature of most of the KS systems we have studied is that their basis tables are ``saturated", i.e., they contain all the orthogonalities of the projectors in them. The basis table is thus completely equivalent to the ``Kochen-Specker diagram" of the projectors (i.e., a graph whose vertices represent the projectors and whose edges join orthogonal pairs of projectors). All the parity proofs associated with a KS system can be obtained simply by dropping selected sets of bases from the basis table.\\

We illustrate the preceding remarks by considering the kite diagram of Fig.1. This diagram gives rise to a system of 32 projectors that form 36 bases (6 pure and 30 hybrid). The most compact parity proof in this system consists of 24 projectors that form 9 bases, which we can denote by the symbol 24-9. However a more informative symbol for this proof is $12^{2}_{2}12^{4}_{2}-4_{4}4_{6}1_{8}$, in which the first half shows the number of projectors of each rank (with the rank indicated as a superscript and the number of occurrences of that projector among the bases as a subscript) and the second half shows the number of bases of each size (with the size indicated as a subscript). Thus the expanded symbol just given implies that this parity proof consists of 12 rank-2 projectors that occur twice each and 12 rank-4 projectors that also occur twice each among 4 bases of size four, four bases of size 6 and one basis of size eight. The kite diagram actually has 33 different types of parity proofs in it, involving all odd numbers of bases from 9 to 17 \cite{WaegellUn}. Each of these proofs has a large number of replicas under symmetry, making the total number of distinct proofs in the system swell to 33152. A similar situation obtains for many of the other KS systems. We should stress that the only parity proofs we consider are ones that are critical, i.e., ones that fail if even a single basis is dropped from them. However, even with this filter, the number of distinct proofs in a KS system typically runs into the thousands. An interesting numerological feature of many of the KS systems we have looked at is that the total number of distinct proofs in them (when all replicas under symmetry are included) is $2^{H}$, where $H$ is half the number of hybrid bases in the system. We have found systems with $H = 9,10$ \cite{Waegell2012}, $12$ (Table 1, left), $20$ (Fig.1, right), $15$ \cite{Waegell2011} and $11,13,16,17$ and plan to give details of them in a future work \cite{WaegellUn}. The higher members of the kite family allow us to construct remarkably compact parity proofs in spaces of high dimension. For example, we have found a proof in 128 dimensions (which can be realized physically using a seven-qubit system) that involves 36 projectors and 9 bases and has the symbol $24^{8}_{2}12^{32}_{2}-4_{4}4_{10}1_{16}$ \cite{WaegellUn}.\\
\section{\label{sec:5}Conclusion}
Our results demonstrate that KS proofs, whether in the observables form advocated by Mermin \cite{Mermin1} or the parity proof form advocated by Peres and others \cite{Cabello1996,Waegell2011,KP1995}, are not rare occurrences but can be assembled in the thousands out of the observables of the Pauli group. A subset of these proofs (namely, those consisting of only a single set of commuting multiqubit observables) can also be converted into GHZ paradoxes. This rich store of proofs is worth exploring for its foundational and practical interest. On the foundational side, it provides new GHZ paradoxes and KS proofs based on small numbers of qubits, which are still the most accessible systems. The experimental realization of these proofs is not trivial because they call for measurements of sets of commuting multiqubit observables in a single context, but this task is not beyond the scope of current or foreseeable technology. Noise and imperfections will of course prevent the realization of the ideal scenarios we have discussed here, but our ``all or nothing" proofs can be converted into Bell inequalities that can be tested experimentally. Our parity proofs are of interest because, following Cabello \cite{Cabello2008}, they can be turned into inequalities that can be used to rule out noncontextuality, and they could also find application in quantum key distribution \cite{BPPeres}, quantum error correction  \cite{Hu2008,Raussendorf2001}, random number generation \cite{Svozil} and parity oblivious transfer \cite{Spekkens}.


\clearpage

\begin{thebibliography}{10}
\providecommand{\url}[1]{{#1}}
\providecommand{\urlprefix}{URL }
\expandafter\ifx\csname urlstyle\endcsname\relax
  \providecommand{\doi}[1]{DOI \discretionary{}{}{}#1}\else
  \providecommand{\doi}{DOI \discretionary{}{}{}\begingroup
  \urlstyle{rm}\Url}\fi

\bibitem{GHZ}
D.M.Greenberger, M.Horne, A.Zeilinger in {\it Bell's Theorem, Quantum Theory and Conceptions of the Universe}, edited by M.Kafatos (Kluwer, Dordrecht, 1989); D.M.Greenberger, M.Horne, A.Zeilinger, A.Shimony {\it Am. J. Phys.} {\bf 58}, 1131 (1990).

\bibitem{Redhead}
Important precursors to Ref \cite{GHZ} were P.Heywood, M.L.G.Redhead, {\it Found. Phys.} \textbf{{\bf 13}}, 481 (1983);
A.Stairs, {\it Phil. Sci.} \textbf{{\bf 50}}, 587 (1983); and
H.R.Brown, G.Svetlichny, {\it Found. Phys.} \textbf{{\bf 20}}, 1379 (1990).

\bibitem{Mermin1}
N.D.~Mermin, {\it Phys. Rev. Lett.} \textbf{{\bf 65}}, 3373 (1990); {\it Rev. Mod. Phys.} \textbf{{\bf 65}}, 803 (1993).

\bibitem{Mermin2}
N.D.~Mermin, {\it Phys. Rev. Lett.} \textbf{{\bf 65}}, 1838 (1990).

\bibitem{Pagonis}
C.Pagonis, M.L.G.Redhead, R.K.Clifton, {\it Phys. Lett. A} \textbf{{\bf 155}}, 441 (1991).

\bibitem{DiV}
D.P.DiVincenzo, A.Peres, {\it Phys. Rev. A} \textbf{{\bf 55}}, 4089 (1997).

\bibitem{Zuk}
D.Kaszlikowski, M.Zukowski, {\it Phys. Rev. A} \textbf{{\bf 66}}, 042107 (2002).

\bibitem{Cerf}
N.J.Cerf, S.Massar, S.Pironio {\it Phys. Rev. Lett.} \textbf{{\bf 89}}, 080402 (2002). Eq.(18) of this paper describes a GHZ paradox for an arbitrary even number of qudits, but which fails for $d = 2$ (the case of qubits).

\bibitem{Lee}
J.Lee, S.W.Lee, M.S.Kim {\it Phys. Rev. A} \textbf{{\bf 73}}, 032316 (2006).

\bibitem{Tang}
W.Tang, S.Yu, C.H.Oh, arXiv:1206.2718.

\bibitem{Cabello2001}
A.~Cabello, {\it Phys. Rev. Lett.}  \textbf{{\bf 86}}, 1911 (2001); ibid. \textbf{{\bf 87}}, 010403 (2001).

\bibitem{Aravind2002}
P.K.~Aravind, {\it Found. Phys. Lett.} \textbf{{\bf 15}}, 399 (2002).

\bibitem{Note1}
GHZ paradoxes based on four qubits were presented in Refs. \cite{Cabello2001} and \cite{Aravind2002}. Ref. \cite{Aravind2002} also presents a proof for $2n$ qubits, where $n$ is an odd integer greater than or equal to 3, but this does not cover the case of $8,12,16,\cdots$ qubits. Another difference between Ref. \cite{Aravind2002} and the present work is that the former uses $n$ Bell states shared symmetrically between two observers (with each observer getting one member of every Bell pair) whereas the present work uses a multiparticle entangled state of $n$ qubits shared among $n$ observers. The present entangled states have a variety of interesting properties not possessed by the former states, as discussed in the text.

\bibitem{Note2}
The justification for this statement is as follows. We used a computer program to construct all sets of five mutually commuting observables for a system of four qubits whose product is the negative identity and which also has the property that each single-qubit observable occurs an even number of times among the five observables (since only such sets can give rise to a GHZ paradox). When we did this we found that all the sets satisfying these conditions were equivalent to the one in Table 1, left, under a permutation of the qubits and/or the observables X,Y,Z.

\bibitem{Briegel}
H.J.Briegel, R.Raussendorf, {\it Phys. Rev. Lett.} \textbf{{\bf 86}}, 910 (2001).

\bibitem{Yeo}
Y.Yeo, W.K.Chua, {\it Phys. Rev. Lett.}  \textbf{{\bf 96}}, 060502 (2006); C.Wu, Y.Yeo, L.C.Kwek, C.H.Oh, arXiv:quant-ph/0611172.

\bibitem{Borras}
A.Borras, A.R.Plastino, J.Batle, C.Zander, M.Casas, A.Plastino {\it J. Phys. A: Math. Gen.}  \textbf{{\bf 38}}, 1119 (2005).

\bibitem{Choudhury}
S.Choudhury, S.Muralidharan, P.K.Panigrahi, {\it J. Phys. A: Math. Theor.} {\bf 42}, 115303 (2009).

\bibitem{Cabello1996}
A.Cabello, J.M.Esteberanz, G.Garcia-Alcaine, {\it Phys. Lett. A}  \textbf{{\bf 212}}, 183 (1996).

\bibitem{Waegell2011}
M.Waegell and P.K.Aravind, {\it J. Phys. A: Math. Theor.} {\bf 44}, 505303 (2011); M.Waegell, P.K.Aravind, N.D.Megill, M.~Pavi{\v c}i{\'c}: {\it Found. Phys.} \textbf{{\bf 41}}, 883 (2011).

\bibitem{KP1995}
M.~Kernaghan, A. Peres, {\it Phys. Lett. A} \textbf{{\bf 198}}, 1 (1995); A.Peres, {\it J. Phys. A} \textbf{{\bf 24}}, L175 (1991).

\bibitem{Cabello2008}
A.Cabello, {\it Phys. Rev. Lett.} \textbf{{\bf 101}}, 210401 (2008); P.Badziag, I.Bengtsson, A.Cabello, I.Pitowsky, ibid. \textbf{{\bf 103}}, 050401 (2009).

\bibitem{BPPeres}
H.~Bechmann{-P}asquinucci, A.~Peres: {\it Phys. Rev. Lett.} \textbf{{\bf 85}},
  3313 (2000); K. Svozil: ArXiv:0903.0231 (2009);  A. Cabello, V.DAmbrosio, E.Nagali, F.Sciarrino, {\it Phys. Rev.} \textbf{{\bf A84}}, 030302(R) (2011).

\bibitem{Hu2008}
D.Hu, W.Tang, M.Zhao, Q.Chen, S.Yu, C.H.Oh {\it Phys. Rev. A}  \textbf{{\bf 78}}, 012306 (2008).

\bibitem{Raussendorf2001}
R.Raussendorf, H.J.Briegel, {\it Phys. Rev. Lett.} \textbf{{\bf 86}}, 5188 (2001).

\bibitem{Svozil}
K.~Svozil: {\it Phys. Rev. A} \textbf{{\bf 79}}, 054306 (2009).

\bibitem{Spekkens}
R.W.~Spekkens, D.H. Buzacott, A.J.Keehn, B.Toner, G.J.Pryde, {\it Phys. Rev. Lett.} \textbf{{\bf 102}}, 010401 (2009).

\bibitem{WaegellUn}
M.Waegell, P.K.Aravind, in preparation.

\bibitem{Waegell2012}
M.Waegell, P.K.Aravind, {\it J. Phys. A: Math. Theor.} \textbf{{\bf 45}}, 405301 (2012).

\end{thebibliography}

\end{document}